\documentclass[aps,preprint,prb,showpacs,amsmath,superscriptaddress,onecolumn,floatfix,nobalancelastpage]{revtex4-1}

\usepackage{amsmath}
\usepackage{bm}
\usepackage{graphicx}
\usepackage{grffile} 
\usepackage{color}
\usepackage{esint}
\usepackage{textpos}
\usepackage[usenames,dvipsnames]{xcolor}
\usepackage[section]{placeins}

\newcommand{\ketbra}[2]{\left|#1\middle>\middle<#2\right|}

\begin{document}


\author{M.~A.~Sentef}
\affiliation{Stanford Institute for Materials and Energy Sciences (SIMES),
Stanford University and SLAC National Accelerator Laboratory, Menlo Park, CA 94025, USA}
\affiliation{HISKP University of Bonn, 53115 Bonn, Germany}
\email{michael.sentef@hiskp.uni-bonn.de}
\author{M.~Claassen}
\affiliation{Geballe Laboratory for Advanced Materials, Stanford University, Stanford, CA 94305, USA}
\author{A.~F. Kemper}
\affiliation{Lawrence Berkeley National Lab, 1 Cyclotron Road, Berkeley, CA 94720, USA}
\author{B.~Moritz}
\affiliation{Stanford Institute for Materials and Energy Sciences (SIMES),
Stanford University and SLAC National Accelerator Laboratory, Menlo Park, CA 94025, USA}
\affiliation{Department of Physics and Astrophysics, University of North Dakota, Grand Forks, ND 58202, USA}
\author{T.~Oka}
\affiliation{Department of Applied Physics, University of Tokyo, Tokyo 113-8656, Japan}
\author{J.~K.~Freericks}
\affiliation{Department of Physics, Georgetown University, Washington, DC 20057, USA}
\author{T.~P.~Devereaux}
\affiliation{Stanford Institute for Materials and Energy Sciences (SIMES),
Stanford University and SLAC National Accelerator Laboratory, Menlo Park, CA 94025, USA}
\affiliation{Geballe Laboratory for Advanced Materials, Stanford University, Stanford, CA 94305, USA}
\email{tpd@stanford.edu}


\title{Theory of pump-probe photoemission in graphene:\\Ultrafast tuning of Floquet bands and local pseudospin textures}

\begin{abstract}
The control of physical properties of solids with short laser pulses is an intriguing prospect of ultrafast materials science. Continuous-wave high-frequency laser driving with circular polarization was predicted to induce a light-matter coupled new state possessing a quasi-static band structure with an energy gap and a quantum Hall effect, coined ``Floquet topological insulator''. Whereas the envisioned Floquet topological insulator requires well separated Floquet bands and therefore high-frequency pumping, a natural follow-up question regards the creation of Floquet-like states in graphene with realistic pump laser pulses. Here we predict that with short low-frequency laser pulses attainable in pump-probe experiments, states with local spectral gaps at the Dirac points and novel pseudospin textures can be achieved in graphene using circular light polarization. We demonstrate that time- and angle-resolved photoemission spectroscopy can track these states by measuring sizeable energy gaps and quasi-Floquet energy bands that form on femtosecond time scales. By analyzing Floquet energy level crossings and snapshots of pseudospin textures near the Dirac points, we identify transitions to new states with optically induced nontrivial changes of sublattice mixing that can lead to Berry curvature corrections of electrical transport and magnetization.
\end{abstract}

\maketitle

The ultrafast optical manipulation of materials by femtosecond laser pulses is rapidly becoming a major guiding theme in condensed matter physics\cite{Fausti14012011,Sobota2012,Caviglia2012}. At the same time, the quest for novel topological states of matter triggered enormous research activity since the discovery of topological insulators\cite{Koenig07}. Merging both of these vibrant fields, a recent work reported the coupling of short laser pulses to surface Dirac fermions in the topological insulator Bi$_2$Se$_3$\cite{Wang13}. This work demonstrated the creation of Floquet-like sidebands during irradiation as well as the opening of a small band gap at the surface state Dirac point for circular light polarization.

Time-reversal symmetry protects massless Dirac fermions on the surface of topological insulators\cite{HasanRMP10,Bernevig06} and, in combination with inversion symmetry, also in graphene in the absence of spin-orbit coupling\cite{NetoRMP}. In a milestone paper, Haldane envisioned that breaking either or both of these symmetries would open a gap at the Dirac points in graphene, allowing
one to tune between a trivial insulator and a Chern insulator\cite{Haldane88}. While equilibrium band gap engineering has become a major theme since the first synthesis of monolayer graphene, it was only recently proposed that circularly polarized, high-frequency laser light could turn trivial equilibrium bands into topological nonequilibrium Floquet bands \cite{Oka09,Kitagawa10,Kitagawa11,Lindner11,Gu11,Calvo11,Dora12,Morell12,RudnerPRX13,IadecolaPRL13,PhysRevB.89.121401,PhysRevLett.112.156801,Fregoso13,Babak,Aditi,Dahlhaus}, coined ``Floquet topological insulator'' (FTI). 

The FTI concept is based on two things: First, in the limit of continuous laser driving at frequency $\Omega$, the temporal periodicity allows one to employ a repeated quasi-energy zone scheme with a temporal Brillouin zone of size $\Omega$. Second, in the high-frequency limit, defined by $\Omega$ being larger than the electronic bandwidth, these repeated zones contain well-separated copies of the original electronic bands spaced by integer multiples $n\Omega$, the so-called Floquet sidebands. The effect of the laser on the original $n=0$ band manifold is perturbative in $1/\Omega$. If the laser is circularly polarized, time-reversal symmetry is broken and an energy gap opens at the Dirac points in graphene due to the fact that photon emission and absorption processes do not commute. The FTI concept then follows from an exact mapping of driven graphene to the Haldane model, leading to a well-defined nonzero Chern number.

Whereas this envisioned high-frequency strong pumping limit that is required for nontrivial topological states is currently experimentally unattainable, a natural follow-up question regards the engineering of local spectral gaps in realistic pump-probe experiments. This leads to the question on which time scales the quasi-steady Floquet regime can be reached when continuous-wave driving is replaced by a short laser pulse. Moreover, it requires the investigation of the low-frequency regime, in which $1/\Omega$ perturbation theory is not applicable. In this regime the overlap of different Floquet sidebands prevents a global topological classification of states. 

In this work, we address these problems by simulating the real-time development of single-particle energy gaps in graphene coupled to short laser pulses, using realistic parameters for time-resolved, angle-resolved photoemission spectroscopy (tr-ARPES). We show that the tr-ARPES band structure shows well-defined Floquet bands provided that a hierarchy of time scales is fulfilled between the duration of the pump pulse, the duration of the probe pulse, and the laser period: $\sigma_{\text{pump}}$ $>$ $\sigma_{\text{probe}}$ $\gg$ $2\pi\hbar/\Omega$. We predict the opening of a Dirac point gap and the formation of Floquet sidebands that form on femtosecond time scales. An important difference to the high-frequency limit arises from the overlap of Floquet sidebands. At frequency dependent critical driving field strengths, we find a sequence of level crossings and energy gap closings at the Dirac points. The analysis of snapshots of pseudospin textures near the Dirac points allows us to identify optically-induced nontrivial changes of sublattice mixing at these level crossing points, that manifest themselves in Berry curvature corrections of electrical transport and magnetization. Even though a global topology cannot be assigned to the low-frequency driven states, we show that the analysis of level crossings and energy gap closings leads to a classification scheme in terms of local gaps and Berry curvatures.

To set the stage for our results, we briefly outline the basic ingredients for the low-energy physics of Haldane's equilibrium model. We start from two Dirac cones with effective Hamiltonian $v_D (q_x \sigma_x \otimes \tau_z + q_y \sigma_y \otimes \bm{I})$. Here $v_D$ is the Dirac point velocity, the Pauli matrices $\bm{\sigma}$ label pseudospin arising from the graphene sublattices $\mathcal{A}$ and $\mathcal{B}$, and $\bm{\tau}$ labels the valley degree of freedom corresponding to the Dirac cones around $K$ and $K'$. Momentum $\bm{q}$ is measured from the respective Dirac points. The pseudospin content $\bm{P}(\bm{q})$ essentially measures orbital band content (see SM for details). For instance, a pseudospin pointing along the $+z$ (``up'') direction means that the band is predominantly of $A$ sublattice character, while a pseudospin pointing along the $-z$ (``down'') direction indicates mainly $B$ sublattice character. Together with the winding of the $P_x$ and $P_y$ ``in-plane'' pseudospin components around the Dirac points, $P_z$ determines the local Berry curvature $\mathcal{F}(\bm{q}) = (\partial_{q_x}\bm{P}(\bm{q})\times\partial_{q_y} \bm{P}(\bm{q}))\cdot \bm{P}(\bm{q})$\cite{TKNN}.

\begin{figure*}[h!t]
	\includegraphics[width=\textwidth]{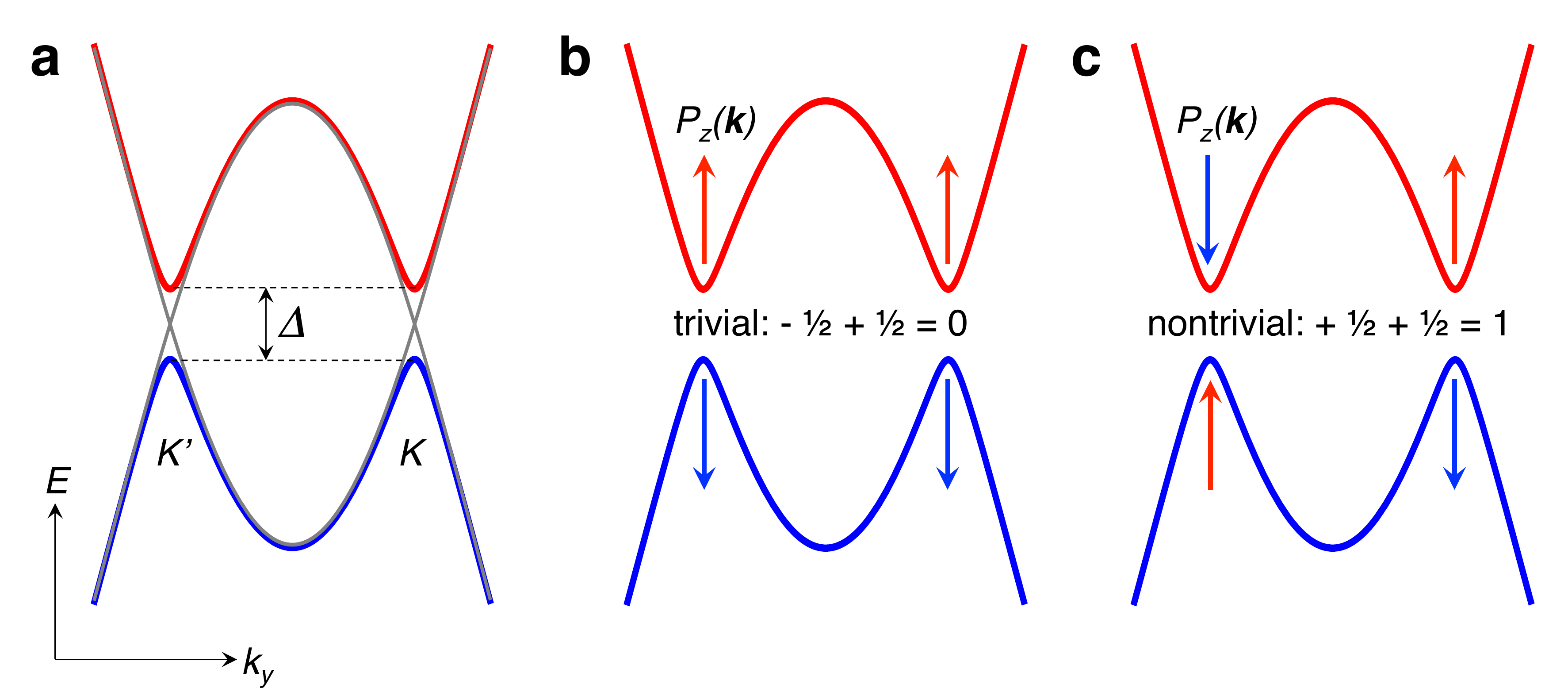}
        \caption{
        {\bf Graphene with band gap.}
        (a) Graphene band structure with a band gap at the Dirac points. (b) Trivial gap structure in Haldane's model: The pseudospin $P_z(\bm{k})$ points in the same direction at $K$ and $K'$. The winding number contributions cancel. This gap structure appears if inversion symmetry is broken, and time-reversal symmetry is intact. (c) Nontrivial gap structure: $P_z(\bm{k})$ points in opposite directions at $K$ and $K'$, leading to a nonzero Chern number $+1$ or $-1$. This gap structure appears if time-reversal symmetry is broken, and inversion symmetry is intact.
        }
        \label{fig1}
\end{figure*}

In Haldane's model, an effective mass term $m \sigma_z \otimes \tau_{\kappa}$ leads to an energy gap $\Delta$ $=$ $2m$ at the Dirac points (Fig.~\ref{fig1}a). Its relative sign between $K$ and $K'$ is determined by $\tau_{\kappa}$ and depends on its origin: If the gap is induced by introducing a staggered sublattice potential breaking inversion symmetry, $\tau_{\kappa}$ $=$ $\tau_0$, implying that the effective mass term has the same sign at $K$ and $K'$, and the out-of-plane pseudospin component $P_z$ is the same at both Dirac points (Fig.~\ref{fig1}b). By contrast, if the gap originates from breaking time-reversal symmetry, $\tau_{\kappa}$ $=$ $\tau_z$, hence $P_z$ points in opposite directions (Fig.~\ref{fig1}c). 

We now come to the discussion of our nonequilibrium results. We start from the minimal honeycomb-lattice tight-binding model of graphene. We drive this system by coupling to a time-dependent, spatially homogeneous electric field modelled as a time-dependent vector potential $\bm{A}(t)$, which couples to the electrons via Peierls substitution. The relativistic magnetic component of the light field is neglected. The pump pulse has a temporal width $\sigma_{\text{pump}}=165$ fs, photon frequency $\Omega=1.5$ eV (laser period of 2.8 fs), with linear or circular light polarization, corresponding to a femtosecond ``pump pulse''. The field strength is given by $A_{\text{max}}$, which is measured in units of the inverse carbon-carbon distance. For graphene, the conversion to the peak electric field strength is $E_{\text{max}} = A_{\text{max}} \times 1060$ mV/$\text{\AA}$ for $\Omega=1.5$ eV. We track the time- and momentum-resolved single-particle spectrum of the pump-driven electrons using a short 26 fs ``probe pulse'' that emits photoelectrons and thereby generates a photocurrent\cite{Freericks09TRARPES,Sentef13}, as measured experimentally with tr-ARPES (see SM). 

\begin{figure*}[h!t]
	\includegraphics[width=\textwidth]{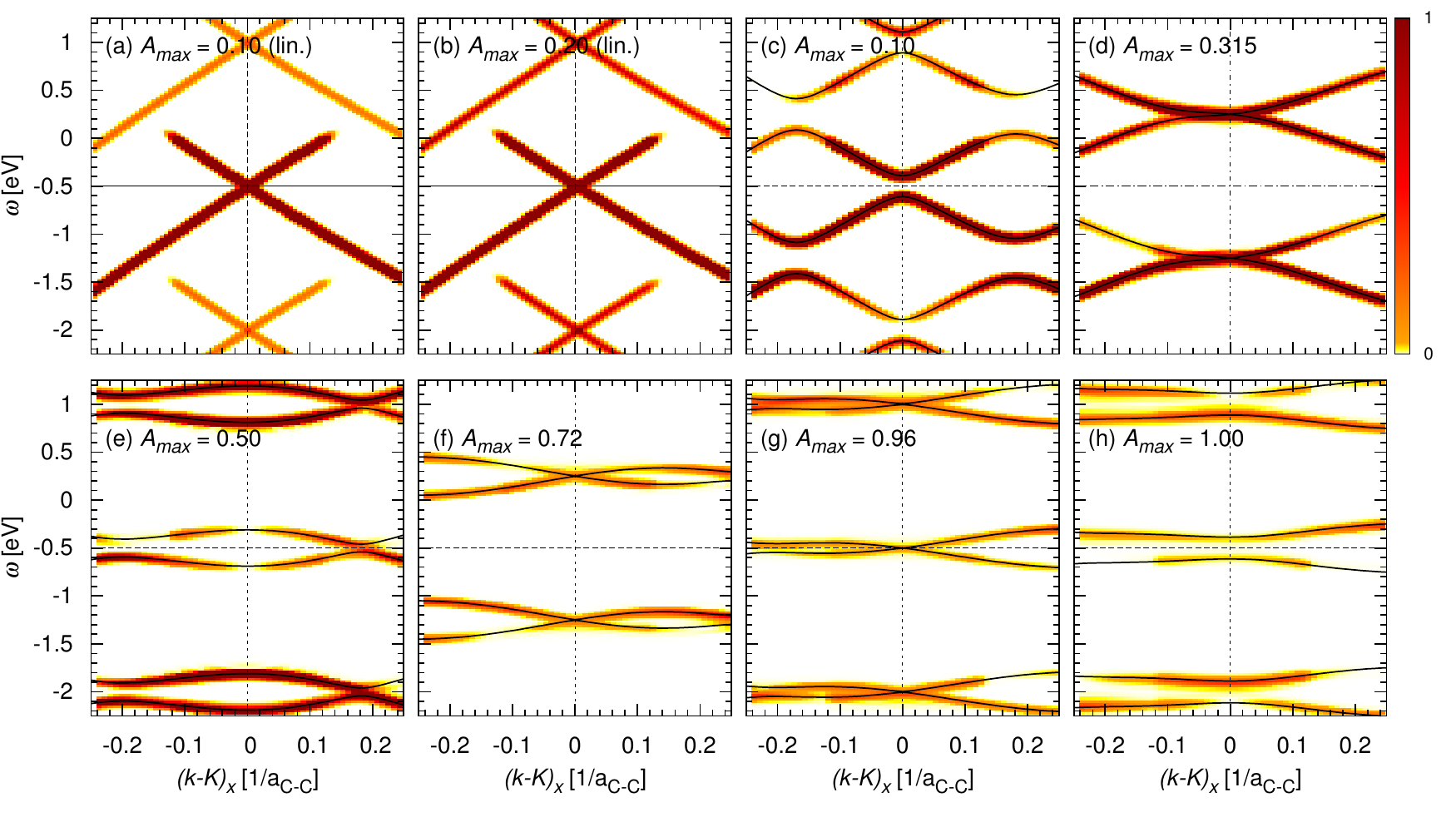}
        \caption{
        {\bf Transient angle-resolved photoemission spectrum.}
        Spectral evolution of irradiated graphene at peak field near the Dirac $K$ point (vertical dashed lines) with Dirac point energy at -0.5 eV (horizontal dashed lines) on a momentum cut along the $\Gamma-K-K'$ direction. The pump laser field is linearly polarized in panels (a) and (b), and circularly polarized in panels (c)-(h). Pump pulse field strengths as indicated. The driving frequency is 1.5 eV. The solid curves show the corresponding quasi-static Floquet band structures. 
        }
        \label{fig2}
\end{figure*}

We first characterize the nonequilibrium band structures using tr-ARPES spectra.  Fig.~\ref{fig2} shows the tr-ARPES spectra on a momentum cut along the $\Gamma-K'-K$ direction near $K$ at peak field ($\Delta t = 0$ fs). We first perform a calculation using pump pulses with linear polarization along the $k_x$ direction for two different field strengths (Fig.~\ref{fig2}(a),(b)). One can see the formation of Floquet sidebands, but since the pump preserves time-reversal symmetry, the spectrum remains gapless at the Dirac point energy. The main effect of the linearly polarized pump is a shift of the Dirac point location from $K$ towards $K'$, which increases with increasing field strength. This Dirac point shift is due to the nonlinearity of bands and does not happen for perfect Dirac cones. 

Next, we turn to circular light polarization, thereby breaking time-reversal symmetry. In Floquet theory, the quasi-static eigenvalue spectrum at finite driving field $A$ shows copies of the original bands shifted by integer multiples of $\Omega$, the so-called Floquet sidebands. Energy gaps of $n$-th order in the field open at avoided level crossings of sidebands which differ by $n$ photon energies. For circular light, an energy gap of second order in the field opens at the Dirac point (see SM). In our tr-ARPES simulation, for a moderate field strength and 1.5 eV photons, an energy gap exceeding 100 meV at $K$ is induced, accompanied by avoided level crossing gaps nearby (Fig.~\ref{fig2}c). Due to the aforementioned hierarchy of time scales, we observe an excellent agreement of tr-ARPES spectra and the quasi-static Floquet band structure obtained by diagonalizing the Floquet Hamiltonian involving large numbers of sidebands (solid lines, see SM). 

At larger field strength (Fig.~\ref{fig2}d), the Floquet bands move closer to each other and cross. They separate again and the Dirac point energy gap decreases (Fig.~\ref{fig2}e). At even larger fields, there is another crossing between Floquet bands (Fig.~\ref{fig2}f) before the Dirac point gap closes (Fig.~\ref{fig2}g) and eventually reopens (Fig.~\ref{fig2}h). At the largest field strength shown here ($A_{\text{max}}$ $=$ 1.00), the bands are almost flat, indicating that the ac Wannier-Stark limit is approached. This creation of flat Wannier-Stark bands in the strong driving limit impedes the continuous growth of the gap with increasing field strength.

\begin{figure*}[h!t]
	\includegraphics[width=0.80\textwidth]{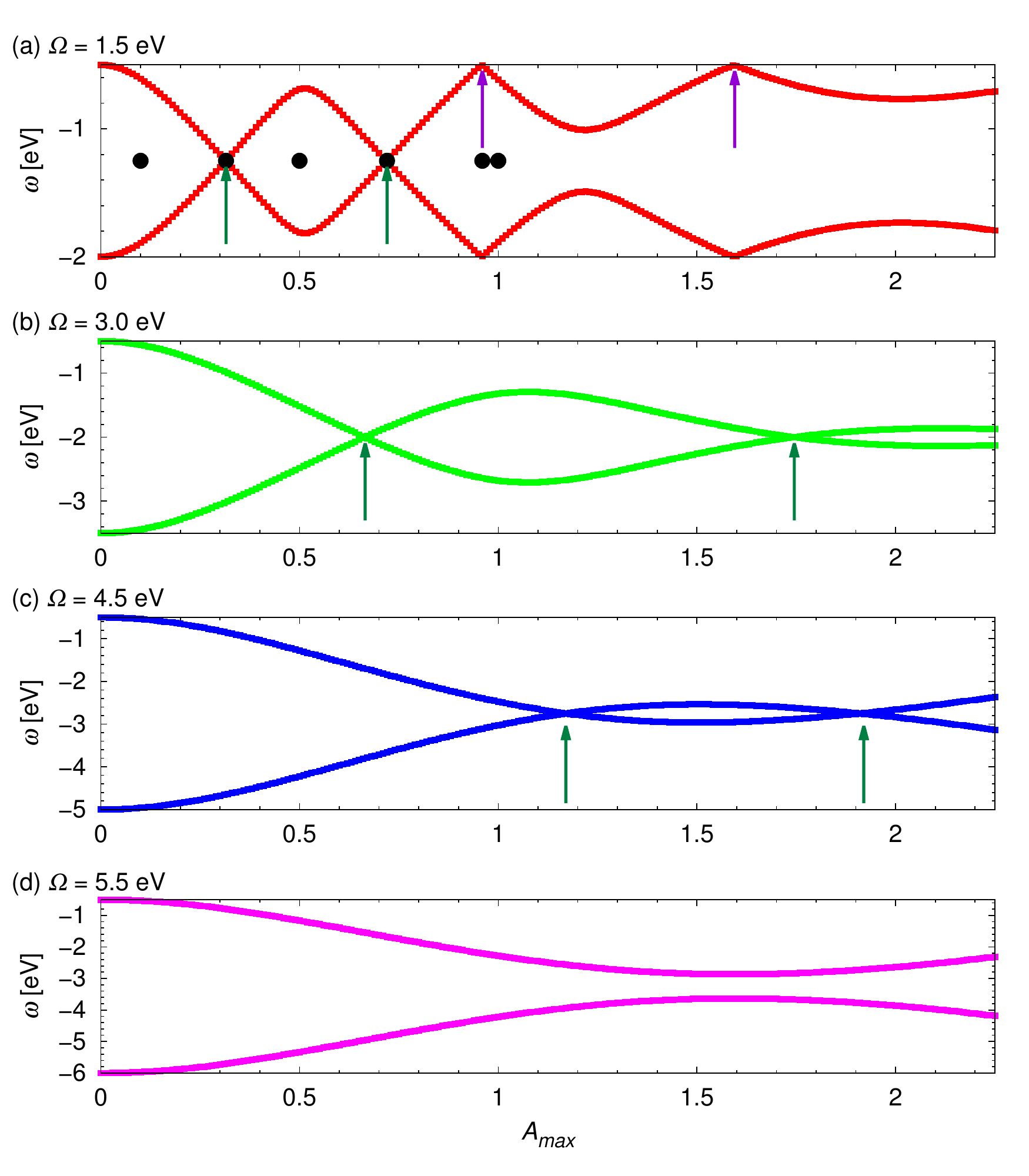}
        \caption{
        {\bf Gap opening and level crossings in Dirac point Floquet spectra.}
        (a)-(d) First two negative Floquet quasi-energies below the Dirac point energy (-0.5 eV) as a function of field strength $A_{\text{max}}$ for driving frequencies as indicated. The crossings between the two Floquet bands as well as the gap closing at higher field strength for $\Omega = 1.5$ eV are indicated by arrows. Black circles in (a) indicate the field strengths used in Fig.~\ref{fig2}(c)-(h).
        }
        \label{fig3}
\end{figure*}

In order to analyze the pump photon frequency dependence of the Dirac point level crossings and gap closing in more detail, we show in Fig.~\ref{fig3} the first two negative Floquet eigenvalues tracking the position of the first two Floquet bands below $E_D$. Fig.~\ref{fig3}a shows the initial gap opening at $\Omega = 1.5$ eV, which is quadratic in the field at small $A_{\text{max}}$, followed by two level crossings between the two Floquet bands indicated by two arrows. The gap at the Dirac point is then closed, indicated by the third arrow. For $\Omega = 3.0$ eV, the field range between the two Floquet band level crossings increases (Fig.~\ref{fig3}b), then decreases at $\Omega = 4.5$ eV, and finally vanishes for $\Omega = 5.5$ eV. The initial quadratic gap opening is the same for all photon frequencies due to the linearity of the graphene bands near the Dirac points. The differences between different photon frequencies at larger fields then arise from the nonlinearity of the bands further away from the Dirac points, which is specific to graphene.

\begin{figure*}[h!t]
	\vspace{-28mm}
	\includegraphics[width=0.88\textwidth]{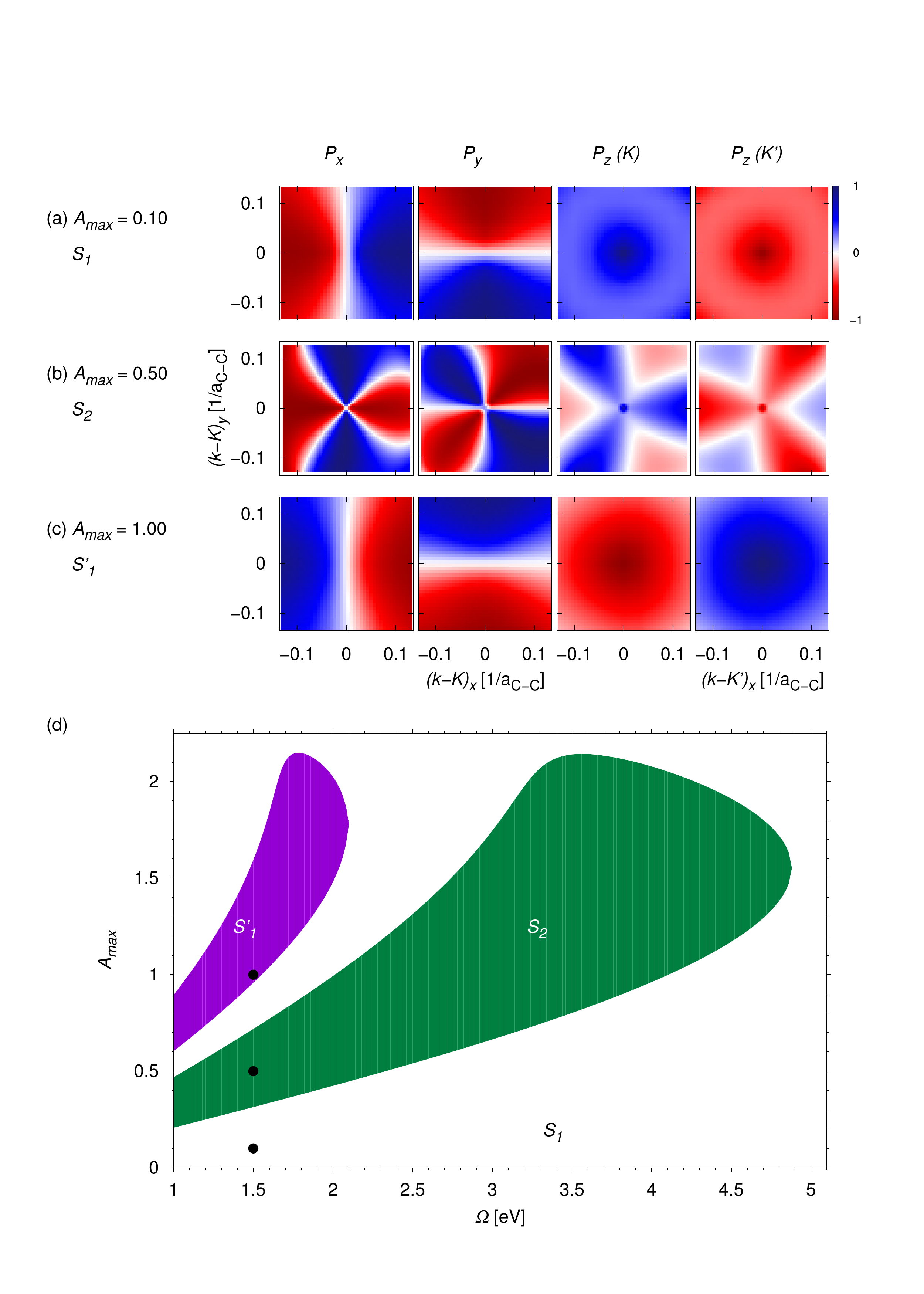}
	\vspace{-12mm}
        \caption{
        {\bf Field-induced changes in local pseudospin textures.}
        (a)-(c) Pseudospin components $P_{x,y,z}(\bm{k})$ near $K$ and $K'$ at peak field (delay time $\Delta t = 0$) for $\Omega = 1.5$ eV and maximal field strengths as indicated. The pseudospin winding changes from $p$-wave in state $S_1$ (a) to four-fold ``nodal lines'' in $S_2$ (b) to $p$-wave in $S'_1$ (c), with opposite pseudospin direction compared to the original $S_1$ state. The $P_z$ component changes sign between $K$ and $K'$ since time-reversal symmetry is broken.
        (d) ``Phase diagram'' of local pseudospin windings near the Dirac points as a function of driving field strength and frequency. The phase boundaries are obtained from the Dirac point level crossings and gap closings (see Fig.~\ref{fig3}). Black circles indicate the parameter values used in panels (a)-(c).
        }
        \label{fig4}
\end{figure*}

We now turn to the discussion of local pseudospin content. Figs.~4(a)-(c) present false color plots of the momentum-resolved pseudospin contents near the Dirac points for the driven system at $\Omega$ $=$ 1.5 eV. At small field before the first level crossing (Fig.~\ref{fig4}a), the $P_x$ and $P_y$ components have two sign changes along a path around $K$, as expected for weakly driven graphene. This nodal structure is directly related to the $q_x \sigma_x + q_y \sigma_y$ term in the effective low-energy Hamiltonian introduced above, which shows that the $P_x$ component transforms like $q_x$ and the $P_y$ component transforms like $q_y$. We coin this state $S_1$, with one ``nodal line'' and therefore a single pseudospin winding in the vicinity of the Dirac points. Importantly, the $P_z$ component changes sign between $K$ and $K'$, which is consistent with breaking time-reversal symmetry. 

When the field is increased through the first level crossing (Fig.~\ref{fig4}b), the character of the local pseudospin textures changes. The effective mass term still changes sign between $K$ and $K'$. Remarkably, the $P_x$ and $P_y$ components double their winding number, changing sign four times along a path around $K$. We call this state $S_2$, since it has two ``nodal lines'' in the vicinity of the Dirac point, which cross at the Dirac point. Although the corresponding ARPES spectrum for the same parameters (Fig.~\ref{fig2}f) has only little spectral weight near $K$ in the Floquet bands close to $E_D$, the pseudospin texture is well-defined at all momenta considered here. Also, other sidebands have higher spectral weight, and each of the Floquet sidebands carries the same pseudospin information.

A similar texture is also obtained for Bernal stacked bilayer graphene\cite{Min08} in a perpendicular electric field\cite{Kumar13}. In this analogy, we stress that the doubled winding in bilayer graphene persists even when the energy gap induced by the electric field goes to zero. By contrast, in our case the doubled winding state $S_2$ vanishes when the Floquet sidebands cross, and gives way to a single winding state $S_1$ in the low-field limit. Also the effective ``perpendicular electric field'' generated by the circularly polarized light pulse in our work points in opposite directions at $K$ and $K'$, in contrast to the static electric field applied in Ref.~\onlinecite{Kumar13}, which corresponds to the Haldane mass term. In any case, the observation of state $S_2$ suggests the possibility to dynamically engineer effective models with higher pseudospin winding numbers, similarly to higher-order spin-orbital textures in topological insulators\cite{Zhang13}. 

When the field is increased further through the second level crossing and the gap closing, one obtains the pseudospin textures shown in Fig.~\ref{fig4}c. Here, all the pseudospin components are flipped compared to the ones in Fig.~\ref{fig4}a, and we coin this flipped state with single pseudospin winding $S'_1$.  

We are now in a position to discuss the ``phase diagram'' of local Berry curvatures, which follow from the pseudospin textures around the Dirac points, as a function of field strength and driving frequency. Fig.~\ref{fig4}d shows the positions of the Floquet band level crossings indicating the transition to a pseudospin texture with a doubled number of ``nodal lines'', as well as the Dirac energy gap closing leading to a state with inverted $P_z$ component. There is an upper frequency limit for the former state in the range of field strengths shown here. This is consistent with the fact that in the infinite-frequency limit, only states with $p$-wave pseudospin textures corresponding to $S_1$ and $S'_1$ were found, which can be understood from the exact mapping to the static Haldane model in this limit. \cite{Cayssol13}

On the one hand, the characterization of nonequilibrium states in terms of local pseudospin textures is restricted to momenta near the Dirac points by Floquet sideband level crossings. Such level crossings generically appear at low driving frequency $\Omega$ because different Floquet sidebands overlap if $\Omega$ is smaller than the electronic bandwidth. On the other hand, sideband level crossings at the Dirac point are the root cause of the appearance of the exotic pseudospin textures in Fig.~4b. The low-frequency behavior of driven graphene is therefore more complicated, but also contains new states which are absent in the high-frequency limit. 

Our combined results show that band gaps induced by breaking time-reversal symmetry in graphene are within reach under realistic experimental conditions. In particular, the achievable energy resolution for probe-photon energies which are sufficiently high to reach the Dirac points\cite{Johannsen13,Gierz13} should allow for the detection of photoemission gap sizes exceeding 100 meV. The change in pseudospin texture near critical driving offers the exciting opportunity of optical manipulation of local Berry curvatures near Dirac points on ultrafast time scales. Moreover, the combination of broken inversion symmetry and broken time-reversal symmetry opens up the possibility of controlling the valley degree of freedom and inducing different energy gaps at the two Dirac points.\cite{Xiao2012,Wu2013,Xu2014,Schwingenschloegl} 

The spectroscopic detection of pseudospin textures requires access to orbital band content. To this end, hexagonal structures with inequivalent orbitals on the $\mathcal{A}$ and $\mathcal{B}$ sublattices having different photoemission probe-energy cross sections could be examined. 
A candidate material for this purpose is hexagonal boron nitride. The demonstration of pseudospin imbalance at the two Dirac points by circularly polarized light in boron nitride would be intriguing. Alternatively, artificial hexagonal lattices with sublattice potentials have already been demonstrated with cold atoms.\cite{Uehlinger13} Thus the proposed pseudospin textures could in principle also be realized in driven ultracold quantum gases.\cite{Tarruell12,Jotzu2014}

%
\newpage
\newpage
\newpage
\newpage



%
\setcounter{figure}{0}
\setcounter{section}{0}
\makeatletter 
\renewcommand{\thefigure}{S\@arabic\c@figure}
\makeatother
\makeatletter 
\renewcommand{\thesection}{S\@arabic\c@section}
\makeatother

\newpage 
\section{Supplementary Material}
\subsection{Methods Summary}
The simulations presented here start from a minimal tight-binding model of spinless electrons with nearest-neighbor hopping on the honeycomb lattice\cite{Wallace47,NetoRMP}. The pump pulse drives the electrons via minimal coupling to a gauge field $\bm{A}(t)$ $=$ $A_{\text{max}} p_{\sigma_{\text{pump}}}(t) \left(\sin(\Omega t) \bm{e}_x + \cos(\Omega t) \bm{e}_y\right)$, with a Gaussian shape function $p_{\sigma_{\text{pump}}}(t)$ $=$ $\exp(-(t-t_p)^2/(2 \sigma_{\text{pump}}^2))$ for a pulse of width $\sigma_{\text{pump}}$ centered around time $t_p$, and photon frequency $\Omega$. The phase shift of $\pi/2$ between the $x$ and $y$ components, represented by the unit vectors $\bm{e}_x$ and $\bm{e}_y$, describes circular light polarization. For comparison we also study linearly polarized light with vanishing $y$ component. Throughout this work we use units where $e = \hbar = c = 1$. The electric field is $\bm{E}(t)$ $=$ $-\partial \bm{A}(t)/\partial t$, and we neglect the relativistic magnetic field of the laser pulse. This means that the electronic spin degree of freedom maintains its full degeneracy, and it is therefore not explicitly included in our calculations. In particular, both the photoemission spectra and the pseudospin textures are identical for both physical spin species.

The time- and angle-resolved photoemission spectroscopy (tr-ARPES) is computed from the trace of the nonequilibrium lesser Green function by a postprocessing step involving the probe-laser-pulse shape $s_W(t)$ with time resolution $W$, which leads to an effective tr-ARPES energy resolution $\propto$ $1/\sigma_{\text{probe}}$\cite{Freericks09TRARPES,Sentef13}. The delay time of a pump peaked at $t_p$ and probe peaked at $t_{pr}$ is given by $\Delta t$ $\equiv$ $t_{pr}-t_p$.

Simulations are performed for an initial equilibrium sample temperature $T$ $=$ 116 K. We typically use 500,000$-$1,500,000 time steps for the computation of the time evolution operator, depending on the field parameters. This corresponds to a maximal time step size of 0.0016 fs. Green functions for tr-ARPES measurements are sampled on a grid with 5,000$-$15,000 real-time steps and a maximal step size of 0.16 fs. The Dirac point velocity is given by $v_D$ $=$ 4.2 eV a$_{\text{C}-\text{C}}$, where a$_{\text{C}-\text{C}}$ $=$ $1.42$ $\text{\AA}$ is the carbon-carbon distance\cite{NetoRMP}. We choose a chemical potential $\mu$ $=$ $0.5$ eV, which sets the Dirac point energy $E_D$ $=$ $-0.5$ eV relative to $\mu$. This choice is motivated by the fact that typical graphene samples on substrates are doped, and that states both below and above the Dirac point energy are occupied in the initial equilibrium states and therefore nicely visible in the tr-ARPES spectra.

The simulation parameters for the pump-probe setup are as follows: The pump laser field has frequency $\Omega$ $=$ 1.5 eV, unless denoted otherwise, implying oscillation periods of 2.58 fs. Its temporal width is $\sigma_{\text{pump}}$ $=$ 165 fs. We vary the peak vector potential $A_{\text{max}}$ $=$ 0.10 $\dots$ 1.00 in units of a$_{\text{C}-\text{C}}$$^{-1}$. This corresponds to peak electric field strengths $E_{\text{max}}$ $=$ $\Omega A_{\text{max}}$ of 106 $\dots$ 1060 mV/$\text{\AA}$ for $\Omega$ $=$ 1.5 eV and the graphene lattice parameters. The photoemission probe pulse has a width $\sigma{\text{probe}}$ $=$ 26 fs. This choice of parameters is motivated by the hierarchy of time scales in the system: The oscillation period for the pump laser light, the temporal width of the probe pulse which controls the time and energy resolution for the tr-ARPES signal, and the temporal width of the pump pulse which controls the nonequilibrium state and ensures a well-defined center frequency for the pump pulse. 

\subsection{Model and time evolution} 
Our goal is to obtain the lesser Green function matrix $\mathcal{G}^<(\bm{k},t,t')$ in $2 \times 2$ orbital space (see below) with matrix elements
\begin{align}
G^<_{\alpha\beta}(\bm{k},t,t') &\equiv \text{i} \langle \alpha^{\dagger}_{\bm{k}}(t) \beta^{}_{\bm{k}}(t') \rangle,
\end{align}
where $ \alpha^{\dagger}_{\bm{k}}$ ($\beta^{}_{\bm{k}}$) is a creation (annihilation) operator for a fermion at momentum $\bm{k}$ in orbital $\alpha$ ($\beta$) $\in \{a,b\}$. As shown below, the photocurrent and pseudospin contents are computed from these lesser Green functions.

Including the field via Peierls substitution, the time-dependent Hamiltonian for $\mathcal{A}$ and $\mathcal{B}$ sublattices with corresponding orbitals $a$ and $b$ reads 
\begin{align}
\mathcal{H}(t) &= \sum_{\bm{k}} \left(\begin{array}{cc} a^{\dagger}_{\bm{k}} & b^{\dagger}_{\bm{k}} \end{array}\right) \left(\begin{array}{cc} 0 & g(\bm{k}-\bm{A}(t)) \\ g^*(\bm{k}-\bm{A}(t)) & 0 \end{array}\right) \left(\begin{array}{c} a^{ }_{\bm{k}} \\ b^{ }_{\bm{k}} \end{array}\right),
\end{align} 
with the Hamiltonian matrix elements 
\begin{align}
g(\bm{k}) &= V\left[2 \cos\left(\frac{\sqrt{3}k_x}{2}\right)\cos\left(\frac{k_y}{2}\right) + \cos(k_y) + \text{i}  \left(-2 \cos\left(\frac{\sqrt{3}k_x}{2}\right)\sin\left(\frac{k_y}{2}\right) + \sin(k_y)  \right) \right],
\label{gk}
\end{align}
where $V$ $=$ 2.8 eV is the nearest-neighbor hopping matrix element matching the graphene bandwidth and Dirac point velocity. In equilibrium, the Hamiltonian has two Dirac points at momenta $K$ and $K'$ given by $(\pm4\pi/(3\sqrt{3}), 0)$ $\approx$ $(\pm 2.4184, 0)$, where momenta and the vector field $\bm{A}(t)$ are measured in multiples of the inverse of the carbon-carbon distance a$_{\text{C}-\text{C}}$.\cite{NetoRMP} 

It is convenient to define the Hamiltonian matrix for momentum $\bm{k}$ in orbital basis, 
\begin{align}
\mathcal{H}(\bm{k},t) &= \left(\begin{array}{cc} 0 & g(\bm{k}-\bm{A}(t)) \\ g^*(\bm{k}-\bm{A}(t)) & 0 \end{array}\right).
\end{align} 
In the absence of a driving field, this Hamiltonian is diagonalized by a rotation $\mathcal{R}(\bm{k})$ at $t$ $=$ $0$, which is a time before the pump pulse is turned on. Note that $t$ $=$ $0$ is used here as a notation for the earliest real time we consider, not to be confused with zero delay time $\Delta t$ $\equiv$ $t_{pr}-t_p = 0$, which refers to the time where the Gaussian pump-pulse envelope is maximal. For later times, the given rotation does \emph{not} diagonalize the Hamiltonian except for accidental cases where the gauge field is an integer multiple of a reciprocal lattice vector. 

The computation of double-time propagators requires the evaluation of the time evolution operators 
\begin{align}
\mathcal{U}(\bm{k},t,t') &= \mathcal{T} \exp\left(-\text{i}\int_{t'}^{t}\mathcal{H}(\bm{k},\bar{t}) \text{d}\bar{t}\right).
\label{time_evolution}
\end{align}
Since $\mathcal{H}(\bm{k},t)$ at different times do not commute with each other, the time ordering $\mathcal{T}$ in $\mathcal{U}(\bm{k},t,t')$ is taken into account by discretization of the real time axis and multiplication of the resulting time-step evolution operators. We then obtain the time evolution operator as $2 \times 2$ matrices in band basis,
 \begin{align}
\mathcal{U}(\bm{k},t,t') &\approx \prod_{j=1}^{N_{t,t'}} \exp\left[-\text{i}\mathcal{H}(\bm{k},t-j\delta t/2)\delta t\right],
\end{align}
where $N_{t,t'}$ is the number of fine time steps of size $\delta t$ between $t$ and $t'$, and the product is understood as time-ordered with later times to the left.

The lesser Green function matrix results from
\begin{align}
\mathcal{G}^<(\bm{k},t,t') &= \mathcal{U}(\bm{k},t,0)^{\dagger} \mathcal{G}^<(\bm{k},0,0) \rangle \mathcal{U}(\bm{k},t',0),\\
\mathcal{G}^<(\bm{k},0,0) &= \mathcal{R}(\bm{k}) \;\text{i}\mathcal{N}(\bm{k}) \mathcal{R}(\bm{k})^{\dagger},
\end{align}
where time ``0'' refers to an initial time where the system is in equilibrium before the pump pulse is turned on, and $\mathcal{N}(\bm{k})$ is the time-independent diagonal matrix of initial equilibrium band occupation with diagonal elements $f(\epsilon_1(\bm{k}))$ and $f(\epsilon_2(\bm{k}))$ corresponding to Fermi function filling for the two energy eigenvalues. 

\subsection{Floquet spectra} 
The Floquet spectra shown in Figures \ref{fig2} and \ref{fig3} are calculated from the Floquet Hamiltonian corresponding to Eq.~(\ref{gk}): 
\begin{align}
	H_F = -\sum_{m\alpha} m\Omega \ketbra{m,\alpha}{m,\alpha} + \sum_{mn} \left[ g_{m-n}(\bm{k}) \ketbra{m,\mathcal{A}}{n,\mathcal{B}} + \textrm{h.c.} \right]
\end{align}
where $g_{m-n}(\bm{k})$ are the Fourier series expansion coefficients of $g(\bm{k} - \bm{A}(t))$:
\begin{align}
	g_{m-n}(\bm{k}) &= \frac{\Omega}{2\pi} \int_0^{\frac{2\pi}{\Omega}} dt~ e^{i(m-n)\Omega t} g(\bm{k} - \bm{A}(t)) \\
		&= \left[ e^{i \left[k_y + \frac{7\pi}{2}(m-n)\right]} + e^{i \left[ \frac{\sqrt{3}}{2} k_x - \frac{1}{2} k_y + \frac{\pi}{6}(m-n) \right] } + e^{-i \left[ \frac{\sqrt{3}}{2} k_x + \frac{1}{2} k_y - \frac{5\pi}{6}(m-n) \right]} \right] J_{m-n}(A)
\end{align}
Here, $J_n(A)$ is the Bessel function of the first kind. As the pump frequencies considered in this work are small with respect to the electronic bandwidth, the corresponding spectrum must be evaluated numerically via truncation of the full Floquet Hamiltonian. In practice, we achieve convergence for $|m| \leq 40$.

\subsection{Time-resolved ARPES formalism}
The computation of the time-resolved photocurrent involves normalized Gaussian probe-pulse shape functions $s_{\sigma_{\text{probe}}}(t)$ of width $\sigma_{\text{probe}}$ centered around time $t$. In the Hamiltonian gauge, the photocurrent (tr-ARPES intensity) at momentum $\bm{k}$, binding energy $\omega$ and pump-probe delay time $\Delta t$ $\equiv$ $t_{pr}-t_p$ is then obtained from\cite{Freericks09TRARPES}
\begin{align}
I(\bm{k},\omega, \Delta t) &= \text{Im} \sum_{a} \int \text{d}t_1 \int \text{d}t_2 s_{\sigma_{\text{probe}}}(t_{pr}-t_1) s_{\sigma_{\text{probe}}}(t_{pr}-t_2) e^{\text{i}\omega(t_1-t_2)} G^{<}_{aa}(\bm{k},t_1,t_2).
\label{ARPES}
\end{align}
In the main text (Fig.~\ref{fig2}) we show tr-ARPES spectra at peak field strength with false color plots of the tr-ARPES intensity $I(\bm{k},\omega,\Delta t=0)$, i.e.~intensity variations as a function of binding energy $\omega$ along selected momentum cuts $\bm{k}$. The location of energy bands $E(\bm{k})$ can be obtained from the maxima in the ARPES intensity as a function of binding energy at constant momentum, the so-called energy distribution curves (EDCs). As seen in Fig.~\ref{fig2} of the main text, these bands are in excellent agreement with quasi-static Floquet bands, whose calculation is described below.

The photocurrent as defined in Eq.~(\ref{ARPES}) is computed from the lesser Green function in a fixed gauge. We would like to point out that this quantity is not gauge-invariant. In fact, the general definition of a gauge-invariant photocurrent that fulfills the positivity criterion $I(\bm{k},\omega, \Delta t)\geq0$ for all $\bm{k},\omega, \Delta t$ in the presence of a field is an outstanding research problem\cite{Freericks14}. The problem likely lies in the neglect of photoemission matrix elements, which can be momentum and field dependent. The photocurrent according to Eq.~(\ref{ARPES}) manifestly fulfills the positivity criterion. In addition, it also matches the Floquet band structure, as shown in the present work. The Floquet band structure is not gauge-invariant either. Importantly, general conclusions drawn from the analysis of Floquet sidebands, level crossings and gap closings are valid even in a fixed-gauge calculation. This is due to the fact that the time-resolved, momentum-integrated photoemission spectrum (tr-PES) is always manifestly gauge-invariant and positive. The tr-PES signal is obtained by integrating the tr-ARPES spectrum in any gauge over all momenta. Hence, conclusions about the presence or absence of energy gaps can be drawn even in a gauge-variant formalism.

\subsection{Pseudospin content} 
We would like to make contact with pseudospin representations of $2 \times 2$ Hamiltonians that reflect their orbital content and fully determine the Berry curvature (see main text). Out of equilibrium and in particular for low driving frequencies, when a description in terms of a simple effective Floquet Hamiltonian is not available, one has to find an analogue of a pseudospin analysis in terms of nonequilibrium Green functions. The pseudospin content of Green functions from our numerical simulations is extracted by expanding the Green function matrices in orbital representation in Pauli matrices,
\begin{align}
G_x(\bm{k},t,t') &\equiv G_{AB}(\bm{k},t,t')+G_{BA}(\bm{k},t,t'),\nonumber\\
G_y(\bm{k},t,t') &\equiv \text{-i}(G_{AB}(\bm{k},t,t')-G_{BA}(\bm{k},t,t')),\nonumber\\
G_z(\bm{k},t,t') &\equiv G_{AA}(\bm{k},t,t')-G_{BB}(\bm{k},t,t').
\end{align} 
The respective pseudospin content $P_{x,y,z}(\bm{k},t)$ specifically for states below the Dirac point energy $E_D$ is obtained by computing the analogue of the tr-ARPES response (\ref{ARPES}) for $G^<_{x,y,z}$,
\begin{align}
P_j(\bm{k},\omega,\Delta t) &= \text{Im} \sum_{a} \int \text{d}t_1 \int \text{d}t_2 s_{\sigma_{\text{probe}}}(t_{pr}-t_1) s_{\sigma_{\text{probe}}}(t_{pr}-t_2) e^{\text{i}\omega(t_1-t_2)} G^<_{j}(\bm{k},t_1,t_2)
\label{pseudospin_spectrum}
\end{align}
with $j \in \{x,y,z\}$. In order to obtain $\bm{P}(\bm{k}, \Delta t=0)$, shown in Fig.~\ref{fig4} of the main text, we integrate (\ref{pseudospin_spectrum}) in a frequency window between $E_D-\Omega/2$ and $E_D$, chosen to represent states below the equilibrium Fermi level in the $n$ $=$ $0$ manifold near the Dirac point, and normalize the resulting vector for each momentum point. We note that the pseudospin behavior is independent of the choice of Floquet sideband for which it is analyzed. Every Floquet band has exactly the same pseudospin content, and the pseudospins in upper and lower bands within a manifold always point in opposite directions. This is easily understood from the fact that at a given momentum the states integrated over all energies contain $\mathcal{A}$ and $\mathcal{B}$ orbital content alike.

We note that in contrast to the Berry curvature obtained from pseudospin representations, the orientation of pseudospin vectors is basis-dependent. For the lattice problem, in-plane pseudospins can be rotated by moving the origin in the real-space unit cell, whereby the Hamiltonian matrix elements $g(\bm{k})$ acquire a momentum-dependent phase factor. However, our conclusions about the relative orientations of pseudospins between Dirac points and the structure of the $S_1$, $S_2$ and $S'_1$ states are independent of this basis choice. 

\subsection{Evolution of spectra and pseudospin across the $S_2$ phase}
\begin{figure*}[h!t]
	\includegraphics[width=\textwidth]{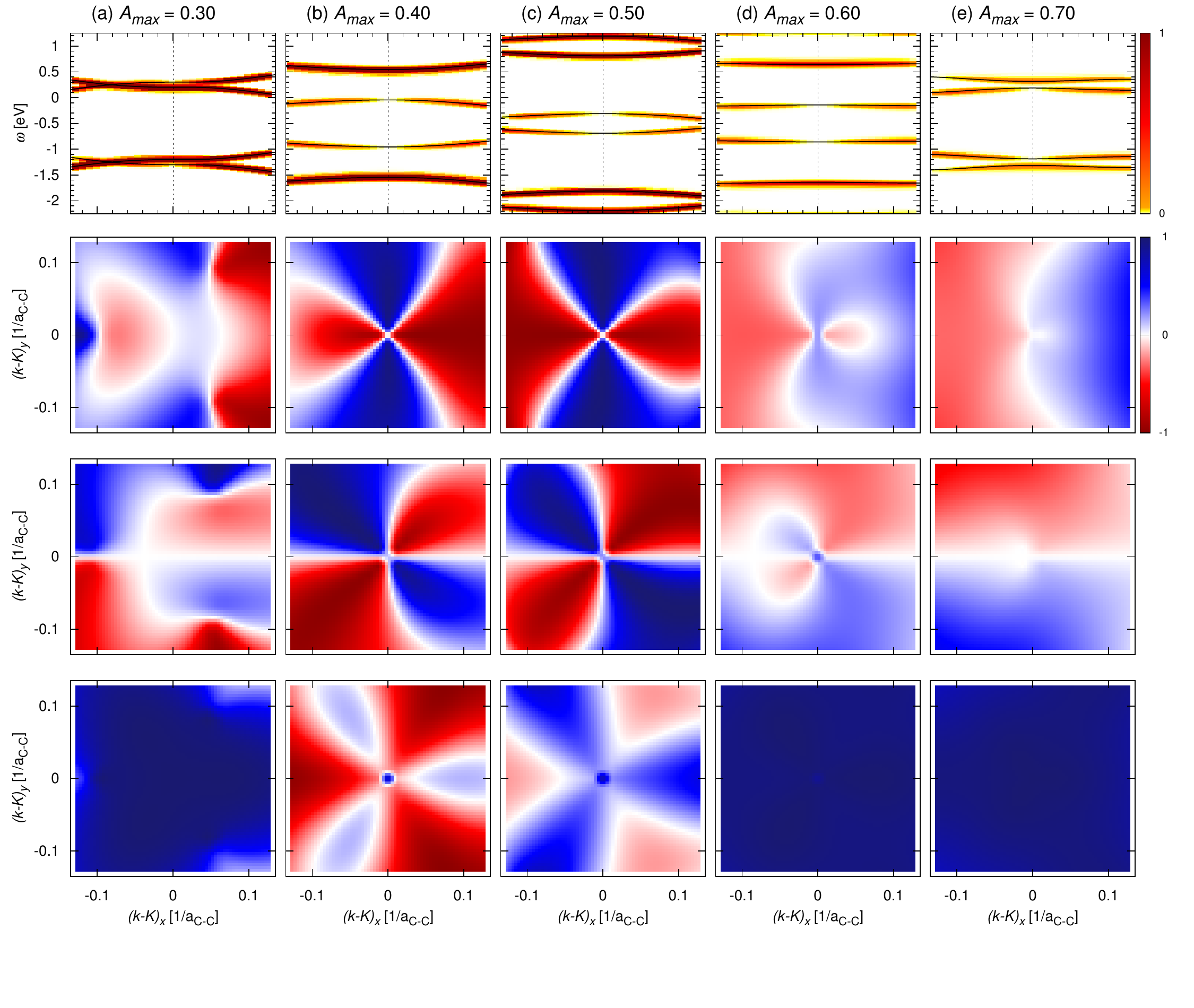}
        \caption{
        {\bf Pseudospin texture evolution around $S_2$ phase for 1.5 eV pulse}
        (a)-(e) ARPES spectrum and pseudospin components $P_{x,y,z}(\bm{k})$ near $K$ for different field strengths as indicated.
        }
        \label{figs1}
\end{figure*}

\begin{figure*}[h!t]
	\includegraphics[width=\textwidth]{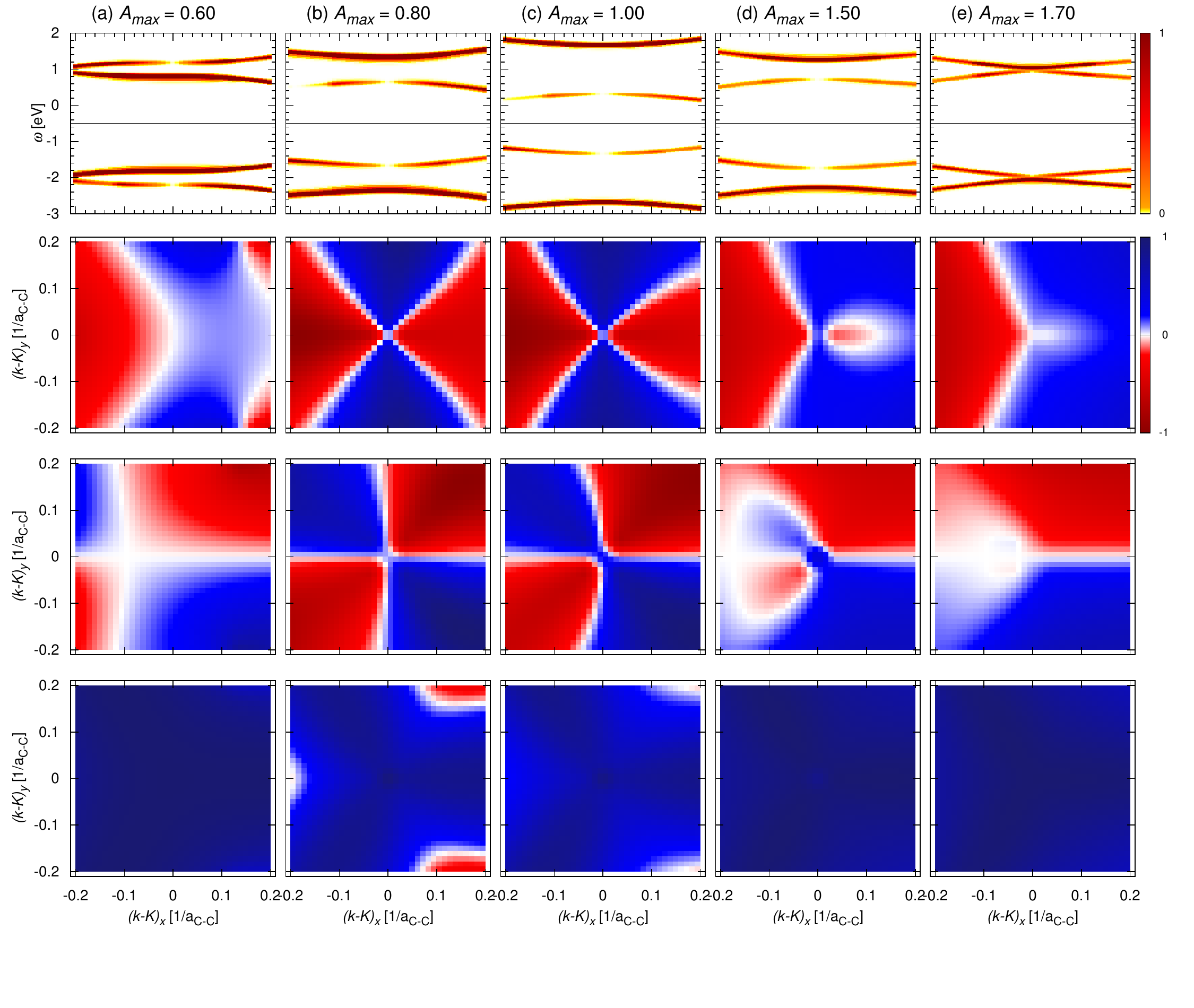}
        \caption{
        {\bf Pseudospin texture evolution around $S_2$ phase for 3.0 eV pulse}
        (a)-(e) ARPES spectrum and pseudospin components $P_{x,y,z}(\bm{k})$ near $K$ for different field strengths as indicated.
        }
        \label{figs2}
\end{figure*}

\begin{figure*}[h!t]
	\includegraphics[width=\textwidth]{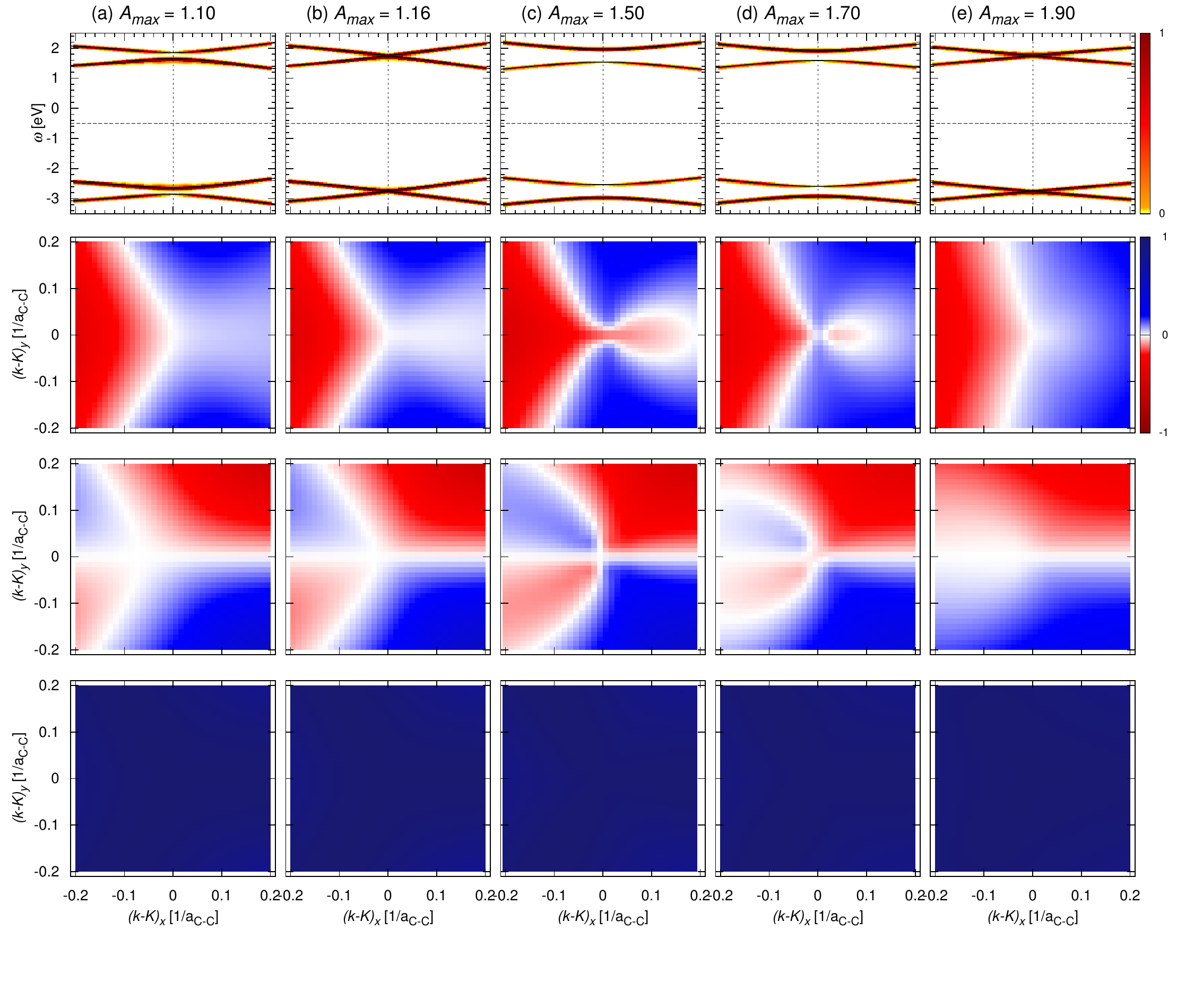}
        \caption{
        {\bf Pseudospin texture evolution around $S_2$ phase for 4.5 eV pulse}
        (a)-(e) ARPES spectrum and pseudospin components $P_{x,y,z}(\bm{k})$ near $K$ for different field strengths as indicated.
        }
        \label{figs3}
\end{figure*}

Figures \ref{figs1}, \ref{figs2}, and \ref{figs3} show how the $S_2$ state emerges from the $S_1$ state for different driving frequencies $\Omega =$ 1.5, 3.0, and 4.5 eV. The upper panels show the ARPES spectra for field strengths near the $S_2$ phase (see Fig.~\ref{fig4}d). At the first sideband level crossing which characterizes the lower $S_2$ phase boundary, additional nodes appear in the $P_x$ and $P_y$ pseudospin components, leading to a four-fold nodal structure. At the second sideband level crossing, these additional nodes disappear, and the nodal structure goes back to $p$-wave. This characterizes the upper boundary of the $S_2$ phase.

\newpage
\bibliography{bib}{}
\bibliographystyle{naturemag}

\newpage
{\bf Acknowledgments.} We acknowledge helpful discussions with Patrick Kirchmann, Sri Raghu, Xiao-Liang Qi, Shou-Cheng Zhang, and Bruce Normand. This work was supported by the Department of Energy, Office of Basic Energy Sciences, Division of Materials Sciences and Engineering (DMSE) under Contract Nos. DE-AC02-76SF00515 (Stanford/SIMES), DE-FG02-08ER46542 (Georgetown), and DE-SC0007091 (for the collaboration). Computational resources were provided by the National Energy Research Scientific Computing Center supported by the Department of Energy, Office of Science, under Contract No. DE- AC02-05CH11231. J.K.F. was also supported by the McDevitt bequest at Georgetown. A.F.K. was supported by the Laboratory Directed Research and Development Program of Lawrence Berkeley National Laboratory under U.S. Department of Energy Contract No. DE-AC02-05CH11231.


\newpage

\newpage

\newpage

\end{document}